\documentclass{aa}
\def\beq{\begin{equation}} \def\eeq{\end{equation}}
\def\etal{{\em et al.}}\def\mean#1{{\vphantom{\tilde#1}\bar#1}}
\def\LCDM{$\Lambda$CDM}\def\h{\,h^{-1}}\def\hm{\h\hbox{Mpc}}\def\dd{{\rm d}}
\def\ns#1{_{\rm #1}}\def\av{a\ns{v}}\def\aw{a\ns{w}}\def\ds{\dd s}
\def\Vav{{\cal V}}\def\fw{{f\ns w}}\def\fv{{f\ns v}}\def\ab{{\bar a}}
\def\fvi{{f\ns{vi}}}\def\fwi{{f\ns{wi}}}\def\QQ{{\cal Q}}\def\abn{\ab_{0}}
\def\rhb{\mean\rho}\def\OM{\mean\Omega}\def\bH{\mean H}\def\gb{\mean\gamma}
\def\Hv{H\ns v}\def\Hw{H\ns w}
\def\OMM{\OM_M}\def\OMR{\OM_R}\def\OMk{\OM_k}\def\OMQ{\OM_{\QQ}}
\def\tw{\tau}\def\etw{\eta\ns w}\def\tH{\widetilde{H}}\def\fvn{f_{{\rm v}0}}
\def\etb{\mean\eta}\def\rw{r\ns{w}}\def\Deriv#1#2#3{{#1#3\over#1#2}}
\def\rhc{\rhb\ns{cr}}\def\dOM{\dd{\Omega_2}^2}\def\fvf{\left(1-\fv\right)}
\def\X#1{_{\lower2pt\hbox{$\scriptscriptstyle#1$}}} \def\pt{\partial}\def\Hh{H}
\def\w#1{\,\hbox{#1}}\def\kms{\w{km}\w{s}^{-1}}\def\kmsMpc{\kms\w{Mpc}^{-1}}
\def\goesas{\mathop{\sim}\limits}\def\frn#1#2{{\textstyle{#1\over#2}}}
\def\Fi{\hbox{\scriptsize\it fi}}\def\ffi{f_{\Fi}}
\usepackage{txfonts}
\begin{document}
\title{Comment on ``Hubble flow variations as a test for inhomogeneous cosmology''}
\author{David L.~Wiltshire}
\institute{School of Physical and Chemical Sciences,
       University of Canterbury, Private Bag 4800, Christchurch 8140,
       New Zealand\\
       \email{David.Wiltshire@canterbury.ac.nz}}

\date{Received 12 December 2018 / Accepted 12 February 2019}

\abstract{\citet{SMKW18} have performed a novel observational test of the
local expansion of the Universe for the standard cosmology as compared to an
alternative model with differential cosmic expansion. Their analysis employs
mock galaxy samples from the Millennium Simulation, a Newtonian $N$--body
simulation on a \LCDM\ background. For
the differential expansion case the simulation has been deformed in an attempt
to incorporate features of a
particular inhomogeneous cosmology: the timescape model. It is shown that
key geometrical features of the timescape cosmology have been
omitted in this rescaling. Consequently, the differential expansion model
tested by \citet{SMKW18} cannot be considered to approximate the
timescape cosmology.}

\catcode`\@=11
\def\aa@manuscriptname{%
  {{\bf624} (2019) A12
  \hspace{\stretch{1}}%
  \copyright\ ESO 2019
}} \def\today{\href{https://doi.org/10.1051/0004-6361/201834833}
{https://doi.org/10.1051/0004-6361/201834833}}
\def\aa@numarticle{A\&A 624, A12}
\def\aa@textidlineempty{E-print version: arXiv:1812.01586 [astro-ph.CO]}
\setlength{\aa@afterheadboxskip}{12 mm}
\catcode`\@=12
\maketitle

\section{Introduction}

The standard Lambda Cold Dark Matter (\LCDM) cosmology is built on the
assumption that average cosmic expansion exactly follows that of a
Friedmann-Lema\^{\i}tre-Robertson-Walker (FLRW) model, and that all
deviations from uniform expansion are described by peculiar velocities,
which can be expressed exactly in terms of local Lorentz boosts
about the FLRW background. However, this is not true of general
inhomogeneous cosmological solutions in general relativity, nor in any theory
that incorporates key principles of general relativity.

Any observational test of differential cosmic expansion is therefore
an important probe of the foundations of the standard cosmology. Given
the high degree of isotropy of the cosmic microwave background, a
notion of average isotropic expansion does apply on the large scale
(though not necessarily given by the FLRW model). Tests of differential
expansion must therefore be performed on scales comparable to that
over which an average isotopic expansion is seen to emerge, namely
scales of order at least $70$--$120\hm$ \citep{Hogg05,Scrimgeour12}, and
ideally extending to a few times this scale.

Tests of differential cosmic expansion on such scales rely on very large
catalogues of galaxy, group and cluster distances and redshifts, which are
noisy and are subject to numerous observational biases which must be
accounted for. Furthermore, any tests are ideally performed in a model
independent manner, which also requires removing assumptions of the FLRW
model which are often taken for granted in many analyses. To date, such a model
independent test has been performed for full sky spherical averages of local
expansion \citep{WSMW13,MW16}, using the COMPOSITE \citep{WFH1,WFH2} and
Cosmicflows-II \citep{CT12} catalogues. It was found with very strong
Bayesian evidence that the spherically averaged expansion is significantly
more uniform in the rest frame
of the Local Group (LG) of galaxies than in the standard CMB rest frame
\citep{WSMW13}. However, while this may at first seem at odds with the
expectations of the standard cosmology, it was subsequently shown by
\citet{KS16} that the result is consistent with Newtonian $N$-body simulations
in the \LCDM\ framework, given a suitably large bulk flow.

In a new paper, \citet{SMKW18} rigorously perform a new type of
test of differential expansion that they have previously proposed
\citep{SMW12}. They consider line-of-sight averages
that account for intervening structures on each line of sight, and
possible effects on the variation of expansion. This is
a considerably more ambitious test than the previous tests involving spherical
averages, as it requires detailed knowledge of the intervening structures
on any line of sight sampled. This can compound any problems relating
to observational and statistical biases.

\cite{SMKW18} analyse fundamental plane distances \citep{SMZC13,SBM15} which
they combine with information from the SDSS \citep{sdss12} and 2MRS
\citep{2mass} surveys to create a catalogue of structures in the local
Universe covering some 22.7\% of the northern hemisphere sky. Setting aside
possible systematic uncertainties arising from incomplete sky coverage, which
\citet{SMKW18} discuss, then the test that they propose based on
determining the fraction of ``finite infinity regions'' \citep{fit1,clocks}
along individual lines of sight is a perfectly reasonable one, given robust
model-independent estimates of the masses of all galaxies along the lines
of sight.

In a magnitude-limited survey, and with large intrinsic scatter in the
data, model-independent estimates of the masses of all galaxies close to
the line of sight are impossible. Consequently, \citet{SMKW18} choose
to estimate both the masses and systematic uncertainties in the
case of the \LCDM\ model in a combined analysis that includes mock galaxy
catalogues from the Millennium Simulation \citep{mill05}. While certainly
justified in the case of the \LCDM\ model, any rescaling of the simulation
to attempt to mimic non-FLRW inhomogeneous expansion has inherent problems. In
addition to various systematic issues discussed by \citet{SMKW18},
I will point out a further geometrical issue that they have not
considered.

\section{Key geometrical features of the timescape cosmology}

The timescape model \citep{clocks,sol,obs} is a phenomenological
cosmology model without dark energy, which provides one possible
interpretation of the \citet{B00,B01} averaging scheme for general
inhomogeneous cosmological solutions in general relativity. In this
framework Einstein's equations apply exactly on small scales on which
the fluid approximation for the average energy momentum tensor holds.
However, average evolution on larger scales need not follow an exact
solution of Einstein's equations. In particular, it need not coincide with
a FLRW model with constant spatial curvature on large scales. A
dynamical coupling of matter and geometry on small scales which allows
spatial curvature to vary is a natural feature of general relativity. The
requirement that spatial curvature remains constant on arbitrarily
large scales of cosmological
averaging is {\em not} a natural consequence of any principles of general
relativity. Rather the FLRW models are historically the best known and tested
means of imposing average spatial homogeneity and isotropy
on the largest scales, to be consistent with observations, albeit
with the introduction of dark energy.

Since generic inhomogeneous cosmologies can exhibit arbitrarily large
differential expansion, any proposal to describe average cosmic evolution
which differs substantially from the FLRW model must explain why average
cosmic expansion nonetheless appears to be so close to homogeneous and
isotropic. An interpretative framework is also required for the
Buchert averaging scheme, since it deals with statistical volume averages
and does not automatically incorporate a means to relate local
observables to the statistical quantities.

In the timescape model both of these matters are dealt with by
revisiting Einstein's strong equivalence principle, and extending
it to general averages of the cosmological Einstein equations
\citep{cep}. In the presence of gradients of spatial curvature between
expanding regions of vastly different densities, the regional
Hubble parameter related to the quasilocal\footnote{The word ``local''
as typically used in phrases such as ``the local Universe'' is ambiguous as it implies a
choice of scale. In general relativity, ``local'' strictly means the
neighbourhood of a point over which gravity can be neglected -- scales much
smaller than galaxies. As soon as one deals with larger regions in which gravity
cannot be neglected then another terminology -- quasilocal -- is required.}
expansion is calibrated in terms of regional rulers and clocks. But the relative
calibration can vary from region to region.

The observation of average
spatial homogeneity is then accounted for differently. As a consequence of
the cosmological equivalence principle \citep{cep} it is recognized
that expansion appears to be uniform because the actual quasilocal
expansion {\em is} uniform in terms of a canonical choice of regional rulers
and clocks that varies from region to region. In a universe which
grows to be dominated in volume by (negatively curved) voids at late
epochs, there is a systematic drift between the
volume-average rulers and clocks (that best describe average cosmic
evolution) and the rulers and clocks of ideal observers in overdense regions
where the mass of the universe is mostly concentrated.
Implementing this requires care.

In the ``two phase'' model that has been
studied to date \citep{clocks,sol,obs,dnw} the {\em average volume},
$\Vav=\Vav\ns i\ab^3$, expands as a disjoint union of spatially flat ``walls''
and negatively curved voids. The walls are formally a union of the ``finite
infinity regions'' \citep{clocks}, which are the compact boundaries enclosing
all gravitationally bound structures within which the density averages to
the timescape model critical density. The volume-average scale factor, $\ab$, is
related to the regional scale factors $\aw$ and $\av$ of the walls and voids
respectively by
\beq
\ab^3=\fwi\aw^3+\fvi\av^3\ ,\label{bav}
\eeq
where $\fwi$ and $\fvi = 1-\fwi$ represent the fraction of the initial volume,
$\Vav\ns i$, in wall and void regions respectively in the very early universe
when $\fvi\ll1$. One may rewrite (\ref{bav}) as
\beq
\fv(t)+\fw(t)=1,
\eeq
where $\fw(t)=\fwi\aw^3/\ab^3$ is the {\em wall volume fraction} and $\fv(t)=
\fvi\av^3/\ab^3$ is the {\em void volume fraction}. Taking a derivative of
(\ref{bav}) with respect to the Buchert time parameter, $t$, gives
\beq
\bH\equiv{\pt_t\ab\over\ab}=\fw\Hw+\fv\Hv\,,\label{bareH}
\eeq
where $\Hw\equiv(\pt_t\aw)/\aw$ and $\Hv\equiv(\pt_t\av)/\av$. This expresses
the relation between the ``bare Hubble parameter'', $\bH$, and effective Hubble
parameters of the walls and voids respectively as determined by
volume-average clocks. This parameterization allows the Buchert evolution
equation \citep{B00} to be written in a form reminiscent of the Friedmann
equation,
\beq
\OMM+\OMR+\OMk+\OMQ=1,\label{beq1}
\eeq
where
$\OMM\equiv\rhb_{M0}\abn^3/(\ab^3\rhc)$,
$\OMR\equiv\rhb_{R0}\abn^4/(\ab^4\rhc)$,
$\OMk\equiv3\alpha^2\fv^{1/3}/(8\pi G\ab^2\rhc)$,
$\OMQ\equiv-(\pt_t\fv)^2/[24\pi G\fv(1-\fv)\rhc]$,
\beq
\rhc\equiv3\bH^2/(8\pi G)\label{density}
\eeq
is the timescape model critical density, $\rhb_{M0}$ and $\rhb_{R0}$ are the
present epoch volume-average matter and radiation densities,
$\alpha^2\equiv-k\ns{v}\fvi^{2/3}c^2$, and $k\ns{v}<0$ is the curvature scale
of the voids.

While the matter and radiation density parameters, $\OMM$ and $\OMR$, scale
with the average volume, $\ab^3$, in a similar manner to
the FLRW model, the spatial curvature fraction, $\OMk$, is very different
since its time-variation depends not only on the average volume, but
also on the fraction of that volume occupied by voids, $\fv$. Finally the
kinematic backreaction term, $\OMQ$, is entirely absent in the FLRW model.
Eq.~(\ref{beq1}) is supplemented by an additional equation for the second
derivative of $\fv$ \citep{clocks,sol,obs}.

Eqs.~(\ref{beq1}), (\ref{density}) refer to averaged geometrical quantities in
terms of a statistically average clock, but apart from the two phase
approximation do not place any restriction on the Buchert scheme.
The timescape model implements the further restriction of the uniform
quasilocal Hubble flow condition as follows. Radial light rays within
finite infinity regions where the metric,
\beq\ds^2_{\Fi}=-c^2\dd\tw^2+\aw^2(\tw)\left[\dd\etw^2+\etw^2\dOM\right]\,,
\label{wgeom}\eeq
is regionally spatial flat, are matched conformally to radial light rays in a
general spherically symmetric metric fit to a solution to the Buchert equations.
This results in an effective {\em dressed} metric on radial lines of sight,
\beq
\ds^2=-c^2\dd\tw^2+a^2(\tw)\left[\dd\etb^2+\rw^2(\etb,\tw)\,\dOM\right]
\label{dgeom}\eeq
where $a\equiv\gb^{-1}\ab$,
\beq\rw=\gb_0\fvf^{1/3}
\int_t^{t\X0}{c\,\dd t'\over\gb(t')(1-\fv(t'))^{1/3}\ab(t')}\,,
\label{eq:rw}\eeq
$\gb_0=\gb(t_0)$, and
\beq\gb\equiv\Deriv\dd\tw{t\ },
\label{clocks1}
\eeq
is the {\em phenomenological lapse function} which gives the difference
of the proper time parameter, $\tw$, of ideal observers in finite infinity
regions (in bound structures like ourselves) from the statistical volume-average time, $t$,
that best encodes average cosmic evolution.

There are two important geometrical issues to note. Firstly, the effective
dressed metric (\ref{dgeom}) is not spatially of constant curvature, and
thus cannot be obtained by a spatial rescaling of a FLRW metric. Secondly,
the {\em dressed Hubble parameter}
\beq\Hh\equiv{1\over a}\Deriv\dd\tw a
={1\over\ab}\Deriv\dd\tw\ab-{1\over\gb}\Deriv\dd\tw\gb
=\gb\bH-\Deriv{\dd}t\gb\,,\label{42}
\eeq
not only has a contribution from rate of change of the average volume,
$\ab^{-1}\pt_\tw\ab$, but also a contribution, $-\gb^{-1}\pt_\tw\gb$,
from the rate of change of the phenomenological lapse function.

As a numerical example, from a fit of the angular scale of the CMB acoustic
peaks using Planck satellite data, \citet{dnw} find a bare Hubble constant,
$\bH_0=50.1\pm1.7\kmsMpc$, and a present epoch lapse
$\gb_0=1.348^{+0.021}_{-0.025}$. The corresponding dressed Hubble constant is
$H_0=61.7\pm3.0\kmsMpc$, which can be understood as being equal to the
fractional
$\tw$-rate of change of the volume-average scale factor, $\bH_{\tw,0}\equiv
\left.{\ab_0}^{-1}\pt_\tw\ab\right|_0=\gb_0\bH_0=67.5^{+3.4}_{-3.5}\kmsMpc$, as
further reduced by $\left.-{\gb_0}^{-1}\pt_\tw\gb\right|_0\goesas-5.8\kmsMpc$.
We note that the nonreduced value $\bH_{\tw,0}$ matches the value of the Hubble
constant obtained for the FLRW model from the same data \citep{Pparm}, albeit
with larger uncertainties since only the angular scale of the acoustic peaks is
fitted.
This is precisely what should be expected since the data fit compares the same
physical scales between last scattering and the present epoch: if one ascribes
this change solely to the rate of change of an average volume by the same clock
then there is a unique result.

\section{Systematic problem of the \citet{SMKW18} rescaling}
To create mock samples for calibrating their differential expansion
model, \citet{SKCMZ16,SMKW18} determine finite infinity radii from
the {\em``millimil''} run of the Millennium simulation as
\beq
R_f=\left(3M\ns{tot}\over4\pi\rho\ns{cr}f\right)^{1/3}\label{Rf}
\eeq
where $M\ns{tot}$ is the mass within a simulated finite infinity region
\citep[Sec.~3.5]{SKCMZ16}, $\rho\ns{cr}$
is the critical density of the simulation, and $f$ is a factor that modifies
the critical density according to
\beq
f=\left(\bH_0\over H_0\right)^2\simeq\left[2(2+\fvn)\over4\fvn^2+\fvn+4\right]
^2\label{Sf}
\eeq
where $\bH_0=\bH(t_0)$ and $H_0=H(t_0)$ are respectively the bare and dressed
Hubble constants for the timescape model as above, $\fvn=\fv(t_0)$, and in the
last equality we have used the exact tracking limit solution \citep{sol,obs}.

Using the above definitions, \citet{SMKW18} empirically define average
fractions of lines of sight in finite infinity regions from the mock catalogues
according to
\beq
\ffi\tH_{0,{\rm w}}+(1-\ffi)\tH_{0,{\rm v}}=\tH_{0,{\rm av}},\label{empirical}
\eeq
assuming $\tH_{0,{\rm av}}$
to coincide with the dressed Hubble constant, $H_0$.
We introduce a tilde on $\tH_{0,{\rm w}}$ and
$\tH_{0,{\rm v}}$ to distinguish these empirical quantities from the
present epoch values of $\Hw$ and $\Hv$ as given in (\ref{bareH}).
Eq.~(\ref{bareH}) is a {\em volume} average using the time parameter, $t$,
whereas (\ref{empirical}) invokes an average over 1-dimensional lines of
sight, with expansion referred to our own time parameter, $\tw$. By the uniform
quasilocal Hubble flow condition, the regionally measured Hubble constant
within spatially flat finite infinity regions would coincide with the bare
Hubble constant, i.e., $\tH_{0,{\rm w}}=\bH_0=50.1\pm1.7\kmsMpc$. However,
eq.~(\ref{42}) which defines the dressed Hubble parameter
is not linearly related to any void fraction. Thus the quantity
$\tH_{0,{\rm v}}$ and the relation (\ref{empirical}) have no obvious
counterparts in the timescape model.

There is, furthermore, a geometrical problem with the volume-based assumption
(\ref{Rf}) that has gone into this construction.
In flat space a sphere of radius, $R$, has Euclidean
volume $V_{\rm E}=\frn43\pi R^3$. For a negatively curved space, the volume of a
sphere of the same fixed radius is {\em larger} than $\frn43\pi R^3$.
Equivalently, a shorter line-of-sight distance is required in a negatively
curved void to obtain the same
volume as the Euclidean case. Regional Hubble parameters are based on volume
expansion. But the relation between line-of-sight distance and volume changes
in the presence of spatial curvature gradients. Consequently, the finite
infinity fraction on a line of sight to any observed galaxy must differ
between the two models as a consequence of geometry.

While the bare Hubble
parameter is related to the critical density $\rhc$ by (\ref{density}), as
shown by eq.~(\ref{42}) the dressed Hubble parameter includes a contribution
from the rate of change of the phenomenological lapse function. Thus it is not
directly related to {\em any} scaling of the average volume as in the
definition of the critical density, $\rho\ns{cr}=3{H_0}^2/(8\pi G)$, in the
FLRW model.

Furthermore, since spatial curvature is not constant in the timescape model,
it is impossible to simply ``correct'' the \citet{SKCMZ16,SMKW18} analysis by
replacing the dressed Hubble constant in (\ref{Sf}) by the term
$\bH_{\tw,0}=\left.{\ab_0}^{-1}\pt_\tw\ab\right|_0$. The underlying problem is
that a density is a mass divided by a volume. For a space of constant spatial
curvature the relation between radius and volume is fixed.
However, in the timescape model the average spatial curvature changes with time
in a way which does not scale in direct proportion to the spatial volume.

In an attempt to overcome some of the limitations of their approach,
\cite{SMKW18} introduce an additional empirical parameter, $b\ns{soft}$,
which can be tuned to adjust the amount of differential expansion. In this way
they can impose a match of the relative Hubble constant to the value
$\bH_0/H_0$ in the limit that the line-of-sight finite infinity fraction
$\ffi\to1$, as predicted by the uniform quasilocal Hubble flow condition.

Unfortunately, even with an additional empirical scaling, a basic problem still
remains. It is impossible for individual galaxies to simultaneously have
identical values of both the relative line-of-sight Hubble parameter
$H_i/H_0$ and of the line-of-sight finite infinity fraction, $\ffi$, as derived
from a simulation by (\ref{Rf}) in both the FLRW and timescape
models.\footnote{This is purely a geometrical statement which holds
irrespective of the fact that the finite infinity notion plays no role in the
Millennium simulation, and irrespective of the fact that the relative
Hubble parameters assume different values of $H_0$.} In the FLRW case
the relationship between the line-of-sight $H_0$ and the average volume of
the simulation is prescribed by Euclidean geometry in any density calibration;
in the timescape case it is
not. For the timescape, the relationship between the dressed parameter $H_0$
and the average volume in the Buchert averages is highly nonlinear, since the
later has negative spatial curvature. The fact that \citet{SMKW18} plot the
expectations for both \LCDM\ and the differential expansion model against
data with identical $H_i/H_0$ and $\ffi$ values indicates that there is an
inconsistency as far as a timescape approximation is concerned.

\section{Discussion and conclusions}

Nonkinematic differential cosmic expansion -- i.e., a distance-redshift
relation which differs from that of a FLRW model plus local Lorentz boosts --
is a generic feature of inhomogeneous cosmological models \citep{bnw}. Such
models include averages of the Einstein equations with backreaction, which may
present viable alternatives to dark energy as a source of late epoch apparent
cosmic acceleration.

\citet{SMKW18} have conceived a novel test for differential expansion, and
have undertaken a heroic effort in their detailed analysis of a large data set.
The data has various systematic limitations when applied to the test in
question.
To overcome this, they combine the data with simulated data
from the \LCDM\ $N$-body Newtonian Millennium simulation \citep{mill05}. This gives a
reasonable estimate of the magnitude of the \citet{kaiser87} effect, which can be
considered as a kinematic differential expansion \citep{bnw}.
They then use a rescaled version of the simulation which accentuates the
differential expansion, since equivalent simulations are unavailable in
the case of the timescape model which inspired their analysis.

Although the details of differential expansion in the timescape model
must differ\footnote{Smaller scale ``fingers of God'' redshift-space
distortions due to velocity dispersions within gravitationally bound
structures are understood in the timescape model similarly to the standard
cosmology. On scales on which space is expanding, however, the conceptual
framework is fundamentally different.} from those of the Kaiser effect in the \LCDM\ model, there are no {\em a priori} grounds
by which we should expect its magnitude to be greater than the Kaiser
effect. Indeed, just as the \LCDM\ model has a restriction on inhomogeneity
-- that average expansion occurs exactly in hypersurfaces of constant spatial
curvature, as in a FLRW model -- the timescape model also has an important
simplifying restriction on inhomogeneity, namely the uniform quasilocal
Hubble flow condition, eq.~(\ref{42}). This restriction is geometrically
very different to that of the FLRW model.

Eq.~(\ref{42}) exactly prescribes how the time rates of change of the
volume-average scale factor and
phenomenological lapse must be combined to match observations when we attempt
to extract a Hubble constant from distance-redshift data in the usual manner
(on scales larger than the statistical homogeneity scale). Any
approximation to the timescape cosmology should incorporate this restriction.
As we have shown here, the \cite{SKCMZ16,SMKW18} scaling does not.

A question remains as to whether some nonlinear deformation of the Millennium
simulation could effectively approximate the restriction of Eq.~(\ref{42}).
This is unclear. \citet{R17} have performed a simulation without dark energy,
the ``AvERA model'', in which standard Newtonian $N$-body codes are evolved
with the Friedmann equations on small scales, and then averaged at each time
step to determine a collective volume-average scale factor in analogy to the
Buchert approach. The resulting distance-redshift relation tracks very close
to that of the timescape model.\footnote{The comoving distance-redshift
relation in the AvERA model \citep[Fig.~5(b)]{R17} is empirically
very close to that of the timescape model fitted to the same Planck data
\citep[Fig.~8(a)]{W14}. The timescape result lies between the \LCDM\ and AvERA
curves, but is closer to the AvERA result overall.} Since it is a Newtonian
$N$-body framework, it does allow for a direct comparison between the
Millennium simulation and a phenomenological backreaction framework \citep{R18}.
However, since the scheme is arrived at by making empirical changes to cosmic
evolution at the level of a computer code, it is unclear how these changes
relate directly to physical questions associated with effective spatial
curvature or simplifying physical principles for average cosmic evolution,
such as those which underlie the timescape model \citep{cep}. Nonetheless, it
may provide a framework for considering the \citet{SMKW18} test, and deserves
further investigation.

Other tests of nonkinematic differential expansion are possible. For example,
apparently anomalous features in the large angle CMB multipoles are a generic
prediction, as quantified by \citet[Eq.~(2.3), (2.4)]{bnw}. Simple
ray--tracing estimates using the Lema\^{\i}tre--Tolman--Bondi model show that
the precision to definitively distinguish nonkinematic differential expansion
from the standard expectation is not reached with present data \citep{dthesis},
however. That test, comparisons of the integrated Sachs--Wolfe effect
\citep{R18}, and the \citet{SMKW18} test are all complementary ways for
testing the possibility of nonkinematic differential expansion once substantial
advances in observational precision are made.

In the case of the timescape cosmology, substantial theoretical advances
are also required to make predictions with the level of detail available
in $N$--body Newtonian simulations on the FLRW background, in
order to allow a direct implementation of the proposed test of \cite{SMKW18}.
\medskip

\noindent {\bf Acknowledgements}\quad I thank Christoph Saulder and Nezihe Uzun
for helpful comments.
%
%

\end{document}